\documentstyle[twoside,fleqn,epsfig,espcrc2]{article}

\newcommand{\AmS}{{\protect\the\textfont2
  A\kern-.1667em\lower.5ex\hbox{M}\kern-.125emS}}
\newcommand{\beq}{\begin{equation}}
\newcommand{\eeq}{\end{equation}}

\title{Interplay of superconductivity with structural phases in a generalized
t-J model}

\author{Roland Zeyher\address{Max-Planck-Institut f\"ur 
Festk\"orperforschung,
70569 Stuttgart, Fed. Rep. of Germany}%
        and 
        Emmanuele Cappelluti\address{Dipartimento di Fisica, 
Universit$\grave{a}$
di Roma I ``La Sapienza'', Piazzale A. Moro, 2, 00185 Roma, Italy}}

\begin{document}

\begin{abstract}
The phase diagram of the $t-J-V$ model is discussed using a $1/N$ expansion
in terms of X operators. It is shown that a flux phase of d-wave symmetry 
is stabilized by the Coulomb interaction at intermediate dopings and 
competes with d-wave superconductivity. Since the flux wave instability
is stronger than the superconducting one optimal doping is essentially
determined by the onset of the flux phase at zero temperature at the doping 
$\delta_c \sim 0.14$ for $J/t=0.3$. 
For $\delta < \delta_c$ the flux phase coexists with 
superconductivity at low and exists as a single phase at higher
temperatures. Due to the different origins of the two instabilities
the flux phase boundary and especially $\delta_c$ is much less sensitive
to impuritiy scattering than the boundary for superconductivity in
agreement with experiments in Zn doped $La-214$ and $(Y,Ca)-123$.
\end{abstract}

\maketitle

\section{INTRODUCTION}

The superconducting transition temperature $T_c$ of many high-$T_c$ 
superconductors can strongly be suppressed by Zn impurities.
Considered as a function of the doping $\delta$ $T_c$ vanishes finally
at a doping $\delta_{QCP}$ which is the $T=0$ endpoint of a 
line separating a metallic phase at higher doping from a pseudogap phase 
at lower doping in the $T-\delta$ plane \cite{Tallon}.
The microscopic nature of the pseudogap phase is presently not clear.
Since its gap has d-wave symmetry like the superconducting gap it has
been suggested that Cooper pairs without long-range phase order 
are responsible for this phase. On the other hand there is
evidence for the occurrence of strong fluctuations in the underdoped
regime due to the proximity of phases different from the superconducting phase.
At zero temperatures these phases may exhibit a
change from long- to short-range order at 
$\delta_{QCP}$ so that $\delta_{QCP}$ is the location of a quantum
critical point. One aim of the present investigation is to study the
$t-J-V$ model at large N to see whether such a quantum critical point
scenario applies in this case and what the relevant additional phase is
besides of the superconducting one. Another aim is to study the relation 
between these two phases.

\section{COMPETITION BETWEEN SUPERCONDUCTIVITY AND FLUX PHASE}

We consider a $t$-$J$-$V$ model with $N$ degrees of freedom per lattice
site on a square lattice. Its Hamiltonian can be written in terms of 
$X$-operators as 
\begin{eqnarray}
H=-\!\sum_{i j  \atop p=1 \ldots N}\!\frac{t_{ij}}{N}X_{i}^
{p 0}X_{j}^{0 p}\!\!&+&\!\!\!\sum_
{i j \atop p,q =1 \ldots N}\!\frac{J_{ij}}{4N}X_{i}^{p q} 
X_{j}^{q p}\nonumber\\
+\sum_{i j \atop p,q =1 \ldots N}
(-\frac{J_{ij}}{4N}
\!\!&+&\!\!\frac{V_{i j}}{2 N}) X_{i}^{p p} X_{j}^{q q}.
\label{htj}
\end{eqnarray}
The internal labels $p$,$q$... consist of a spin label distinguishing
spin up and spin down states and a flavor label counting $N/2$ identical
copies of the original orbital. The coupling constants $t_{ij}$ and $J_{ij}$
are confined to nearest neighbors $i,j$ and simply denoted by $t$ and $J$,
respectively. The X operators satisfy the 
commutation and
anticommutation rules of Hubbard's X operators for all N. 
Moreover, the sum of the diagonal operators is equal to $N/2$ at each site
meaning that only $N/2$ out of the $N$ states per site can be occupied
simultaneously. The first three terms represent the $t$-$J$ Hamiltonian,
the last term a screened Coulomb interaction appropriate for two
dimensions and taken from Ref. \cite{Becca}.
In the following we express all energies in units of $t$ and all lengths in
units of the lattice constant of the square lattice.
The strength of the Coulomb interaction will be characterized by its
value between nearest neighbor sites $V_{nn}$.

The self-energy $\Sigma$ of the one-particle Green's function
is independent of frequency at large N and 
has the general form \cite{Capp}
\beq
\Sigma({\bf k},{\bf q}) = 
\Sigma({\bf k}) N_{c} \delta({\bf q}) +
\phi({\bf k},{\bf q}).
\label{self}
\eeq
$\Sigma({\bf k})$ denotes the self-energy in the normal state.
$\phi$ is an additional contribution describing a new
ground state characterized by the modulation wave vector $\bf q$. For
the determination of phase boundaries $\phi$ can be considered as
infinitesimally small. Calculating explicitly the self-energy from
Eq.(\ref{htj}) one finds 
\beq
\phi({\bf k},{\bf q}) =
\sum_{\alpha} f_{\alpha}({\bf q}) 
F_{\alpha}({\bf k},{\bf q})\:,
\label{ff1}
\eeq
with $\vec{F}({\bf k},{\bf q})=(t({\bf k- q}), 1, J \cos(k_{x}),
J \sin(k_{x})$,
$J\cos(k_{y}),J\sin(k_{y}))$,
and $f_\alpha$ satisfying the homogenous system of equations
\beq
\sum_{\beta}
\left[ \delta_{\alpha\beta} - a_{\alpha\beta}({\bf q})
\right] f_{\beta}({\bf q}) = 0\:.
\label{homo1}
\eeq
The matrix elements $a_{\alpha\beta}({\bf q})$ in 
Eq. (\ref{homo1}) are defined by
\begin{eqnarray}
a_{\alpha\beta}({\bf q}) = 
\frac{1}{N_{c}}\sum_{{\bf p}}
T \sum_{n} 
G_{\alpha}({\bf p},{\bf q})
F_{\beta}({\bf p},{\bf q}) \cdot\nonumber\\
g_{0}({\bf p},i\omega_{n})
g_{0}({\bf p-q},i\omega_{n}),
\end{eqnarray}
with $\vec{G}({\bf k},{\bf q})=(1, t({\bf k}) +V({\bf q})-
\frac{J({\bf q})}{2},-\cos(k_{x})$,
$-\sin(k_{x}),-\cos(k_{y}), -\sin(k_{y}))$,
and $g_0({\bf p},i\omega_n) = 1/(i\omega_n - \Sigma({\bf p})).$
The boundary between the normal state and a possible new phase corresponds
to a nontrivial solution for $f_\beta$ of Eq.(\ref{homo1}). In this way
both the modulation vector $\bf q$ and the components of the order
parameter in the six-dimensional order parameter space is determined.  

Putting first $V=0$ and decreasing doping or temperature from large values
we find for
$J<1$ a first instability of the normal state with respect to the flux phase
order parameter
\beq
\phi_{FL}({\bf k},{P\bf Q}) \sim i(cos(k_x)-cos(k_y)).
\label{orderparameter}
\eeq
\begin{figure}[h]
\vspace*{-2cm}
\centerline{\psfig{figure=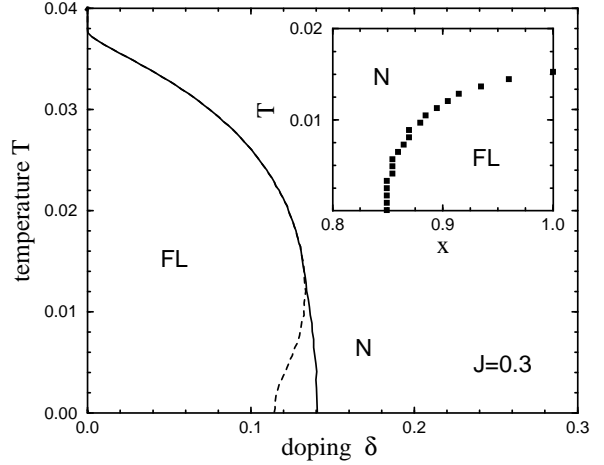,width=20pc,angle=270}}
\vspace*{-1cm}
\caption{Incommensurate (solid line) and commensurate (dashed line) 
phase boundaries in the $T-\delta$ plane. Insert: Instability vector 
${\bf Q} = (1,x) \pi$ as a function of $T$.}
\label{figca3}
\vspace{-0.5cm}
\end{figure}

The solid line in Fig. 1 shows the corresponding instability line
for $J=0.3$ in the $T-\delta$ plane. Writing the modulation vector as
${\bf Q} = \pi (1,x)$ the inset of Fig. 1 shows $x$ as a function of
temperature along the instability line. For $T>0.015$ the flux phase
is commensurate with the wave vector ${\bf Q}_c = \pi (1,1)$, for
$T<0.015$ one wave vector component is incommensurate. Fixing $\bf Q$
to the commensurate value ${\bf Q}_c$ for all temperatures the instability
line is given for $T<0.015$ by the dashed line in Fig. 1 describing
a reentrant behavior of the normal state between $\delta =0.12$ and $0.14$.
This feature is an artefact caused by the assumption of a commensurate
phase at all temperatures. At zero temperature the critical doping
$\delta_c$ is a monotonic function of $J$, with $\delta_c=0$ for $J=0$,
$\delta_c \sim 0.14$ for $J=0.3$ and $\delta_c \sim 0.20$ for $J=0.6$.
Thus $\delta_c$ is near the experimental value for optimal doping and
denotes a quantum critical point because the incommensurate flux phase
has short-range order for any finite temperature if fluctuations are 
taken into account.

\begin{figure}[h]
\vspace*{-1.5cm}
\centerline{\psfig{figure=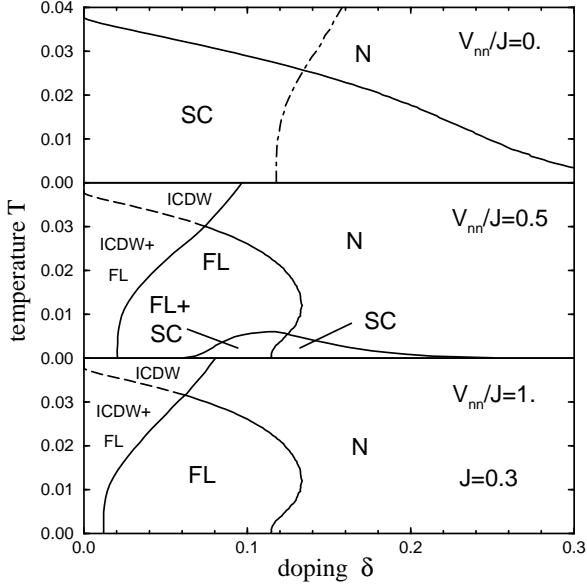,width=20pc,angle=270}}
\vspace*{-0.5cm}
\caption{Phase diagram for $J=0.3$ and different Coulomb 
interaction strengths $V_{nn}$. $N,FL,SC,ICDW$ denote the normal, flux,
superconducting, and CDW phase, respectively, the dot-dashed line in
the upper panel a diverging compressibility in the normal state.}
\label{figca8}
\vspace{-1cm}
\end{figure}

The flux phase is unstable at very low dopings with respect to 
dimers \cite{Erco}. Another 
instability line, indicated by the dot-dashed curve
in the upper panel of Fig. 2, describes a diverging compressibility
of the normal state, i.e., the region to the left of this line would show
phase separation in the normal state. Since the flux phase falls inside
the phase separation regime, the solid line in Fig. 1 has no physical
meaning in the case $V=0$.

The instability line of the normal state towards superconductivity has been
discussed in Ref \cite{Zeyher}. In the BCS-approximation $T_c$ is 
$\sim exp(1/\lambda_i)$, where $\lambda_i$ is the lowest eigenvalue of the 
static kernel of the linearized
gap equation and $i$ denotes one of the 5 representations of the point
group $C_{4v}$ of the square lattice. Only $\lambda_3$ corresponding to
d-wave symmetry exhibits strong negative values. It even diverges at
$\delta_c$ due to critical fluctuations caused by the incipient flux phase.
The numerical solution for $T_c$, which also takes into account retardation,
however, shows that critical fluctuations have only little influence
on $T_c$ and that $T_c$ is mainly determined by the instantaneous 
term. Keeping only this term the solid line in the upper panel of Fig. 2
shows $T_c$ as a function of $\delta$. Interesting is that the
superconducting unlike the normal state is stable against phase
separation. 

Taking also the Coulomb interaction $V$ into account phase separation
is no longer possible but charge density waves (CDW) may be stabilized.
The two lower panels of Fig. 2 show, however, that CDW's may be only
stable at rather low dopings far away from optimal doping and are thus of
less interest for the following. In order to be able to discuss fully the
interplay between superconductivity and the flux phase we treat the
corresponding order parameters in a nonlinear way,
assuming only that the flux phase is commensurate.
The self-consistent equations are then 4x4 matrix equations which
are most conveniently formulated using a Nambu representation with 4 states.

The middle panel of Fig. 2 shows the phase diagram for $V_{nn}/J=0.5$.
For $\delta > \delta_c$ $T_c$ is monotonically increasing with decreasing 
$\delta$ similar as in the upper panel, only its absolute value is
reduced because of the repulsion $V_{nn}$. Below the onset of the flux
phase at $\delta_c$ superconductivity and the flux phase compete with 
each other because both have d-wave symmetry. Since the nesting instability
leading to the flux phase is stronger than the superconducting instability
the flux phase suppresses strongly $T_c$ with decreasing $\delta$ below 
$\delta_c$. As a result $T_c$ assumes a maximum just below $\delta_c$
and decreases both towards lower and higher dopings. Optimal doping is
thus determined at large N by the onset of the flux phase at $\delta_c$.
For $\delta < \delta_c$ the flux and the supercondcuting phase coexist 
for $T < T_c$ whereas the flux phase alone is stable for $T_c < T < T^*$, 
forming
there a pseudogap phase with a d-wave gap in the single particle spectrum.  
The lower panel in Fig. 2 shows the phase digram for $V_{nn}/J=1$.
The Coulomb repulsion is now strong enough to suppress completely $T_c$.
The region of CDW states has been shifted further towards smaller
doping and the flux phase covers a large region to the left of the normal
state.

\section{PHASE DIAGRAM IN THE PRESENCE OF IMPURITY SCATTERING}

Both the flux and the superconducting phase have d-wave symmetry. Their
phase boundaries $T^*$ and $T_c$, respectively, thus should be sensitive
to impurity scattering. Experimentally this holds in the case of Zn 
impurities for $T_c$ but practically not for $T^*$ \cite{Tallon}. 
From a theoretical point
of view $T_c$ and $T^*$ may depend differently on impurity scattering
because the underlying instabilities are quite different. For instance,
nesting effects play an important role in the case of $T^*$ but not in the
case of $T_c$.

\begin{figure}[h]
\vspace*{-0.7cm}
\centerline{\psfig{figure=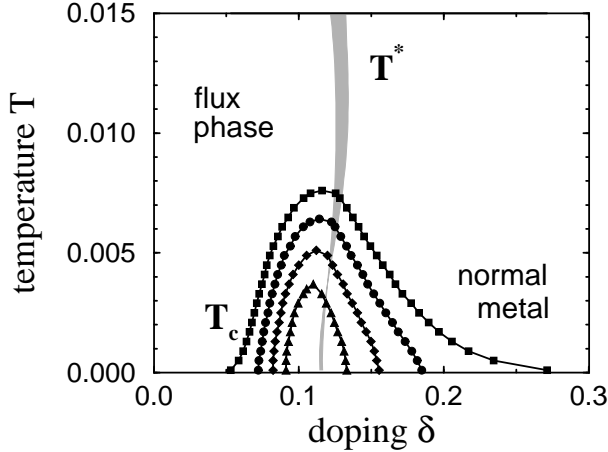,width=20pc,angle=0}}
\vspace*{-1cm}
\caption{Solid lines: Suppression of $T_c$ by impurity scattering, 
calculated with the scattering rates $\Gamma$ =
0 (squares), $2\cdot 10^{-3}$ (circles), 
$4\cdot 10^{-3}$ (diamonds), and 
$6\cdot 10^{-3}$ (triangles). Grey region: Corresponding variation of the
transition temperature $T^*$ to the flux state.}
\label{figg2}
\vspace{-0.5cm}
\end{figure}

In a simple approximation, the effects of impurities in the normal state
can be taken into account by introducing a renormalized 
frequency
\begin{equation}
i\tilde{\omega}_n = i\omega_n  + i \Gamma \frac{\omega_n}{|\omega_n|},
\label{imp1}
\end{equation}
where $\Gamma$ is a scattering rate, here used as a
free parameter proportional to the impurity concentration.
Throughout the flux phase the self-energy due to impurity scattering
is still diagonal in the 4x4 Nambu representation because the
flux order parameter does not couple to the impurities. 
The constant $\Gamma$ in Eq. (\ref{imp1}) accounts in a phenomenological
way both for weak and strong potential scattering by impurities. The
interesting doping region for superconductivity is in our model rather far 
away from the Van Hove singularity, so that the band can be assumed to
be structureless and $\Gamma$ to be independent of frequency.

The
solid lines in Fig. \ref{figg2} show numerical results for $T_c$ as a function
of doping $\delta$ for different scattering rates $\Gamma$, using $J=0.3$
and $V_{nn}=0.5J$. These curves
illustrate the suppression of $T_c$ with increasing scattering rates
$\Gamma = 0$, $2\cdot 10^{-3}$, $4\cdot 10^{-3}$, 
and $6\cdot 10^{-3}$. The corresponding changes in $T^*$, determining the 
phase boundary between the normal state and the flux state, are depicted in 
Fig. \ref{figg2} by the grey region. The chosen values for 
$\Gamma$ correspond roughly to 
$\Gamma \simeq 1.0 T_c$ at optimal doping, and
to $\Gamma \simeq 1.5 T_c$ in the strongly underdoped region, interpolating
between the weak- and the strong-coupling regimes.
One important result of Fig. \ref{figg2} is that the flux phase boundary
$T^*$ is only slightly shifted by impurities, in spite of
the strong suppression of the superconducting critical temperature.
In particular, the critical doping $\delta_c$ at zero temperature 
is almost completely independent of the impurity scattering rate. 
Since in our approach the maximum of $T_c$ as a function of doping 
is essentially determined by $\delta_c$ this means that 
the $T_c(\delta)$ curves shrink to $\delta_c$ with increasing
scattering rate which is a characteristic feature of Fig. \ref{figg2}.
Interpreting Fig. \ref{figg2} in terms of a quantum critical point scenario
means that the corresponding critical doping $\delta_{QCP}$ is given by
$\delta_c$ and that $\delta_{QCP}$ is nearly completely independent of the 
impurity scattering rate. The curves in Fig. \ref{figg2} are in excellent
agreement with the corresponding experimental curves in Zn doped 
(Y,Ca)-123 and La-214, as given in 
Fig. 2 of Ref.\cite{Tallon}.


\begin{thebibliography}{9}
\bibitem{Tallon} J.L. Tallon, C. Bernhard, G.V.M. Williams, and J.W. Loram,
                 Phys. Rev. Lett. 79 (1997) 5294.
\bibitem{Becca}  F. Becca, M. Tarquini, M. Grilli, and C. Di Castro, 
                 Phys. Rev. B {\bf 54} (1996) 12 443.
\bibitem{Capp}   E. Cappelluti and R. Zeyher,
                 Phys. Rev. B {\bf 59} (1999) 6475.
\bibitem{Erco}   E. Ercolessi, P. Pieri, and M. Roncaglia, 
                 Phys. Lett. A 225 (1997) 331. 
\bibitem{Zeyher} R. Zeyher and A. Greco, 
                 Eur. Phys. J. B {\bf 6} (1998) 473.

\end{thebibliography}
\end{document}